\documentclass[a4paper,11pt]{article}
\usepackage{pos}

\DeclareMathOperator{\arcsinh}{arcsinh}
\DeclareMathOperator{\TR}{Tr}
\DeclareMathOperator{\RE}{Re}
\DeclareMathOperator{\IM}{Im}

\title{Dual Polyakov loop model at finite density: phase diagram and screening masses }

\author[a]{Oleg Borisenko}
\author*[b]{Volodymyr Chelnokov}
\author[c]{Emanuele Mendicelli}
\author[d]{Alessandro Papa}

\affiliation[a]{Bogolyubov Institute for Theoretical Physics, National Academy of Sciences of Ukraine,\\
14-b Metrolohichna str. Kyiv, 03143, Ukraine}

\affiliation[b]{Institut für Theoretische Physik, Goethe-Universität Frankfurt, \\
60438 Frankfurt am Main, Germany}

\affiliation[c]{Department of Physics and Astronomy, York University,\\
Toronto, ON, M3J 1P3, Canada}

\affiliation[d]{Dipartimento di Fisica, Universit\`a della
	Calabria, \\
	and Istituto Nazionale di Fisica Nucleare, Gruppo collegato di Cosenza, \\
I-87036 Arcavacata di Rende, Cosenza, Italy}

\emailAdd{oleg@bitp.kiev.ua}
\emailAdd{chelnokov@itp.uni-frankfurt.de}
\emailAdd{emanuelemendicelli@hotmail.it} \emailAdd{papa@fis.unical.it}

\abstract{
We consider a dual representation of an effective
three-dimensional Polyakov loop model for the $SU(3)$ theory 
at nonzero real chemical potential. 
This representation is free of the sign problem and can 
be used for numeric Monte-Carlo simulations. 
These simulations allow us to locate the line of second order
phase transitions, 
that separates the region of first order phase transition 
from the crossover one. 
The behavior of local observables in different phases 
of the model is studied numerically and compared with 
predictions of the mean-field analysis. 
Our dual formulation allows us to study also Polyakov 
loop correlation functions. 
From these results, we extract the screening masses 
and compare them with large-$N$ predictions.
}

\FullConference{%
 The 38th International Symposium on Lattice Field Theory,
 LATTICE2021
  26th-30th July, 2021
  Zoom/Gather@Massachusetts Institute of Technology
}

\begin{document}
\maketitle

\section{Dual effective Polyakov loop model}

The study of the QCD phase structure at nonzero chemical potential
is complicated by the sign problem -- the Boltzmann weight of the
theory becomes complex and thus cannot be treated as a probability
density in the Monte-Carlo simulation.
One of the approaches that can be used to resolve the sign problem
is reformulating the theory in terms of new degrees of freedom,
for which the Boltzmann weight becomes positive, 
enabling numeric simulations. 
While such sign-problem-free formulation is currently unknown 
for full QCD, there has been a number of successful 
dual model constructions for simpler theories \cite{dual-un-staggered,
abelian-color-fluxes,
dual-staggered-qcd,
flux-effective-qcd-polyakov-loop,
su3-flux-simulation,
dual-deconfinement,
dual-polyakov,
flux-spin-su3}.

In this work we consider an effective 3-dimensional 
Polyakov loop model with exact static determinant, 
which can be obtained from a $(3+1)$ $SU(3)$ gauge 
theory coupled with one flavor of staggered fermions 
by going to an anisotropic lattice, 
taking the limit of vanishing space coupling $\beta_s$, 
and performing integration over all spatial gauge fields. 

The model is defined on a 3-dimensional hypercubic lattice
$\Lambda = L^3$, with a partition function 
\begin{equation}
\label{polyakov-pf}
Z(\beta, m, \mu) = \int \prod_x d U(x) \;
\exp \left(\beta \sum_{x,n} \RE \TR U(x) \TR U^\dagger(x+e_n) \right) 
\prod_x B(m, \mu; U(x)) \ ,
\end{equation}
where $U(x)$ are $SU(3)$ variables defined at the site $x$,
corresponding to the Polyakov loops of the original theory,
$\beta$ is an effective gauge coupling constant, 
$m$ and $\mu$ are dimensionless mass and chemical 
potential for the fermion 
($m = m_{\rm ph}/T$, $\mu = \mu_{\rm ph}/T$), and
$B(m, \mu, U)$ is the static fermion determinant
\begin{align}
\label{static-determinant}
    & B(m, \mu, U) = h^{-3} \det \left[ 1 + h e^\mu U \right] \det \left[ 1 + h e^{-\mu} U \right] \ , 
    & h = e^{-N_t \arcsinh a m_{\rm ph}} \approx e^{-m} \ . 
\end{align}

It is clear that at nonzero $\mu$, the static fermion determinant
$B(m, \mu, U)$ is complex, so the Polyakov model in the original formulation exhibits a sign problem. 

The transition to dual variables can be performed by 
Taylor expansion of the gauge part of the partition function ~(\ref{polyakov-pf}), and integrating over the original 
variables $U(x)$ at each site, resulting in the following dual
partition function (see~\cite{dual-polyakov} for a more detailed derivation):
\begin{align} 
\label{dual_PF}
	Z &= h^{-3 L^3} \sum_{\{ r(l) \} = -\infty}^{\infty} 
	\ \sum_{\{ s(l) \}= 0}^{\infty} \ 
	\prod_l  
	\frac{\left ( \frac{\beta}{2} \right )^{|r(l)| + 2s(l)}}{(s(l)+|r(l)|)!
		s(l)!} \ 
	\prod_x R(n(x),p(x))  \ , \\
	\label{nx}
	n(x) &=  \sum_{i=1}^{6}  \left ( s(l_i) + \frac{1}{2} | r(l_i) | \right )
	+ \frac{1}{2} \sum_{\nu=1}^{3} \left ( r_{\nu}(x) - r_{\nu}(x-e_{\nu}) \right )
	\ ,   \\ 
	\label{px}
	p(x) &=  \sum_{i=1}^{6}  \left ( s(l_i) + \frac{1}{2} | r(l_i) | \right )
	- \frac{1}{2} \sum_{\nu=1}^{3} \left ( r_{\nu}(x) - r_{\nu}(x-e_{\nu}) \right )
	\ , \\
	\label{site-weight}
	R(n, p) &= \int dU \; 
	\left(\TR U\right)^n \;
	\left(\TR U^\dagger\right)^p \; 
	B(m, \mu, U) \ .
\end{align}
Here the dual model is defined in terms of two sets of 
variables assigned to the lattice links -- integer 
$r$, and a nonnegative integer $s$. 
The summation over $l_i$ in~(\ref{nx},\ref{px}) 
is done over the six links incident to 
the site $x$.

The integral over the $SU(3)$ invariant measure in~(\ref{site-weight}) can be expressed in terms of simpler group integrals $Q$,
\begin{align}
\label{site-weight-Q}
R(n,p) &= Q(n+1,p) \left ( h_+ + h_-^2 + h_+ h_-^3 + h_+^3 h_-^2  \right ) \\
	& \quad + Q(n,p) \left ( 1+ h_+^3 + h_-^3 + h_+^3 h_-^3  \right )  + 
	Q(n,p+1) \left ( h_- + h_+^2 + h_+^3 h_- + h_+^2 h_-^3  \right )   \nonumber \\ 
	& \quad + Q(n+1,p+1) \left ( h_+ h_- + h_+^2 h_-^2  \right ) + 
	Q(n+2,p) h_+ h_-^2  + Q(n,p+2) h_+^2 h_- \ ,   \nonumber  
	\\
h_{\pm} &= h e^{\pm \mu} \ , \\
Q(n, p) &= \int dU \; \left(\TR U\right)^n \;
	\left(\TR U^\dagger\right)^p \ .
\end{align}

The integrals $Q$ do not depend on simulation parameters and can be precomputed. From the invariance of the integral under multiplication of $U$ by a $Z(3)$ factor follows that $Q(n, p) = 0$ if the difference $(n - p)$ is not divisible by three. In~\cite{sun-integrals} it is shown that for $(n - p) = 3q$,
\begin{equation}
    Q(n,p) \ = \  
	\sum_{\lambda \vdash {\rm min}(n,p)} \ d(\lambda) \ d(\lambda + |q|^3) \ , 
	\label{QSUN}
\end{equation}
where $\lambda = (\lambda_1, \lambda_2, \lambda_3)$ is a partition of $r = {\rm min}(n,p)$:
$\lambda_1 \geq \lambda_2 \geq \lambda_3 \geq 0$, 
$\sum_{i=1}^3 \lambda_i = r$,
$d(\lambda)$ is the dimension of the permutation group $S_r$ in the representation $\lambda$, and $\lambda + |q|^3$ 
denotes a partition $(\lambda_1 + |q|, \lambda_2  + |q|, \lambda_3  + |q|)$. One sees that $Q(p + 3q, p)$ is a positive integer. Thus, at nonzero $h$, $R(n, p) > 0$ for 
any arguments $n, p$ showing that the partition function~(\ref{dual_PF}) is indeed free from the sign problem (at 
$h=0$ the partition function still does not exhibit a sign problem, though the triality condition $n - p = 3 q$ results 
in a condition on the $s$ and $r$ values, that has to be satisfied for the configuration to have a nonzero weight).

A definition of energy density, baryon number density and quark condensate in terms of dual degrees of freedom is easy to obtain by differentiating the partition function~(\ref{dual_PF}) with respect to $\beta$, $\mu$, and $m$, correspondingly,
\begin{align}
    E &= \frac{1}{3 L^3} \frac{\partial \ln Z}{\partial \beta} = 
    \frac{2}{3 \beta L^3} 
    \sum_l \left\langle 2 s(l) + |r(l)| \right\rangle \ , \\
    B &= \frac{1}{L^3} \frac{\partial \ln Z}{\partial \mu}
    =  \frac{1}{L^3}
     \sum_l \left\langle \frac{\partial \ln R(n(x), p(x))}{\partial \mu} \right\rangle \ , \\
     Q &= \frac{1}{L^3} \frac{\partial \ln Z}{\partial m}
    =  \frac{1}{L^3}
     \sum_l \left\langle \frac{\partial \ln h^{-3} R(n(x), p(x))}{\partial m} \right\rangle \ , \\
\end{align}

Introducing sources for $\TR U$, $\TR U^\dagger$ in~(\ref{polyakov-pf}) before the integration, allows one 
to define the expectation values of Polyakov loop products as
expectation values of the ratios of $R$ functions. For example
\begin{align}
    \left\langle \TR U(x) \right\rangle &=
    \left\langle \frac{R(n(x) + 1, p(x))}{R(n(x), p(x))} \right\rangle \ , \\
    \left\langle \TR U(x)^\dagger \right\rangle &=
    \left\langle \frac{R(n(x), p(x) + 1)}{R(n(x), p(x))} \right\rangle \ , \\
    \left\langle \TR U(x) U(y)^\dagger \right\rangle &=
    \left\langle \frac{R(n(x) + 1, p(x)) R(n(y), p(y) + 1)}{R(n(x), p(x)) R(n(y), p(y))} \right\rangle \ .
\end{align}
Note that calculation of these observables in an actual simulation relies heavily on $R$ being nonzero for every value of 
$n, p \geq 0$, so this simple approach cannot be used for
$h = 0$.

\section{Phase structure and local observables}

To simulate numerically the partition function~(\ref{dual_PF}) we implemented a simple Metropolis update algorithm that attempts small updates to the $s$ and $r$ link variables. 
We performed the simulations on lattices with $L = 10, 16, 24$, and (in a few cases) $32$, making $2\cdot 10^5$ thermalization updates, followed by $10^6$ measurements with $50$ updates between measurements. The number of Metropolis hits $n_{hit}$ in multihit update varied between 5 and 2000, while keeping the acceptance rate $p_{acc} \approx 0.5$.

\begin{figure}
        \centering
        \includegraphics[scale=0.2]{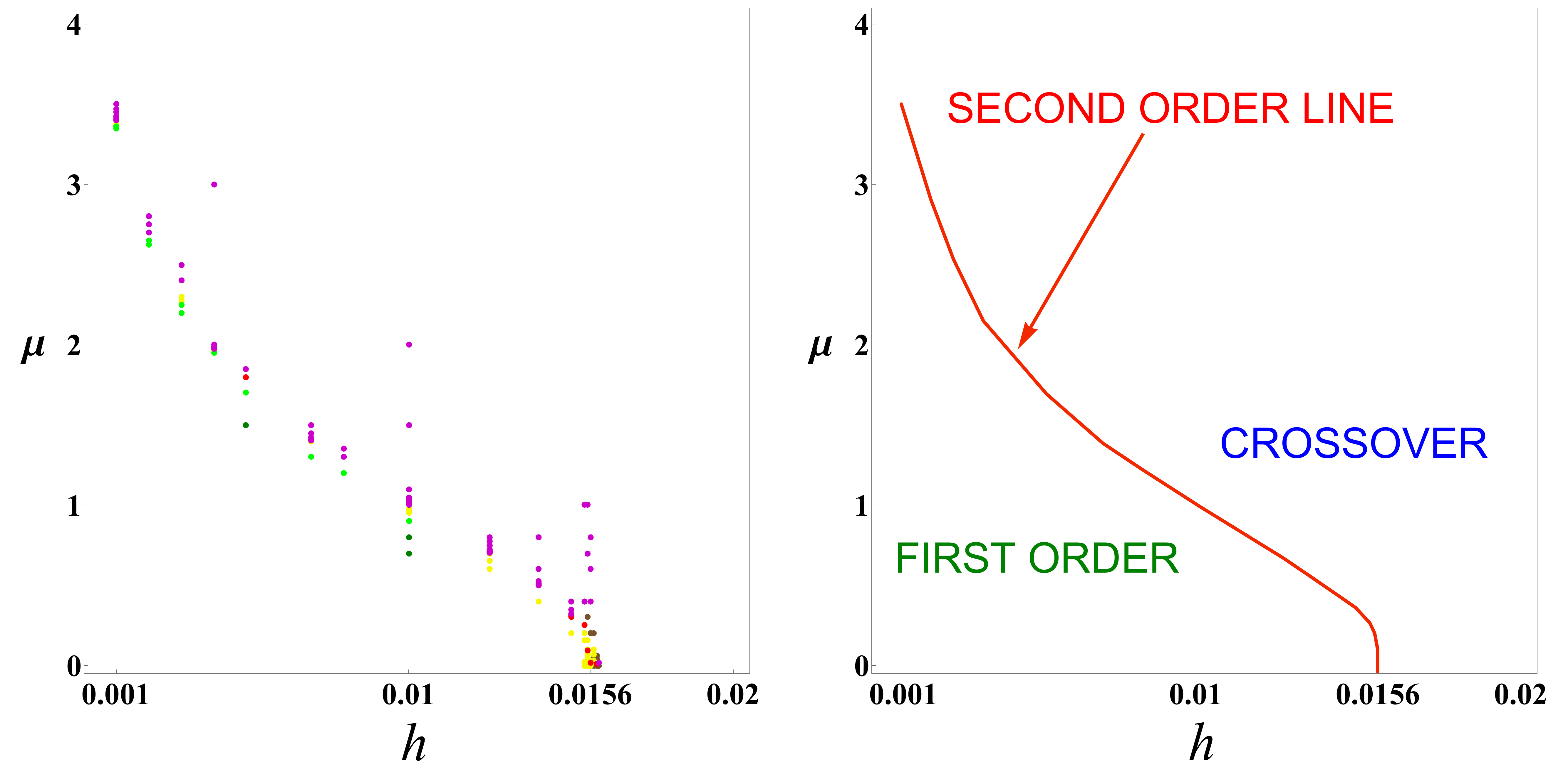}
        \caption{Phase diagram of the Polyakov loop model: (left) simulated points (colored according to the value of $\gamma/\nu$, see~\cite{local-observables}), (right) the interpolated phase diagram}
        \label{fig:phase-diagram}
\end{figure}

For small values of $h$ the Polyakov loop model exhibits a deconfinement phase transition. To study the order of the transition we calculated the ratio of the peak values of the magnetic susceptibility (for fixed $h$ and $\mu$, and varying $\beta$ -- the peak value appears at $\beta_{pc}(L)$),
\begin{equation}
 \chi_M = L^2 \left(
 \left\langle \left(\TR U(x)\right)^2 \right\rangle - 
 \left\langle \TR U(x) \right\rangle^2 \right) \ ,
\end{equation}
and extracted the critical index ratio $\frac{\gamma}{\nu}$
from the results for two lattice sizes
\begin{equation}
    \left(\frac{\gamma}{\nu}\right)_{L_1, L_2} = 
    \frac{\ln {\chi_M}_{\rm peak}(L_2) - \ln {\chi_M}_{\rm peak}(L_1)}
    {\ln L_2 - \ln L_1} \ ,
\end{equation}
and compared it to $1.9638(8)$ -- the ratio for the $3d$ 
Ising model~\cite{critical-indices}. The results are collected in Fig.~\ref{fig:phase-diagram}. As we see for $\mu = 0$
the second order phase transition happens at $h_2(0) \approx 0.0156$, for $h < h_2(0)$ the transition is of first order, and for $h > h_2(0)$ we have a crossover. Introducing a nonzero chemical potential $\mu$ leaves the picture qualitatively the same, but reduces $h_2$: $h_2(\mu) < h_2(0)$.

\begin{figure}
\centering
\includegraphics[scale=0.08]{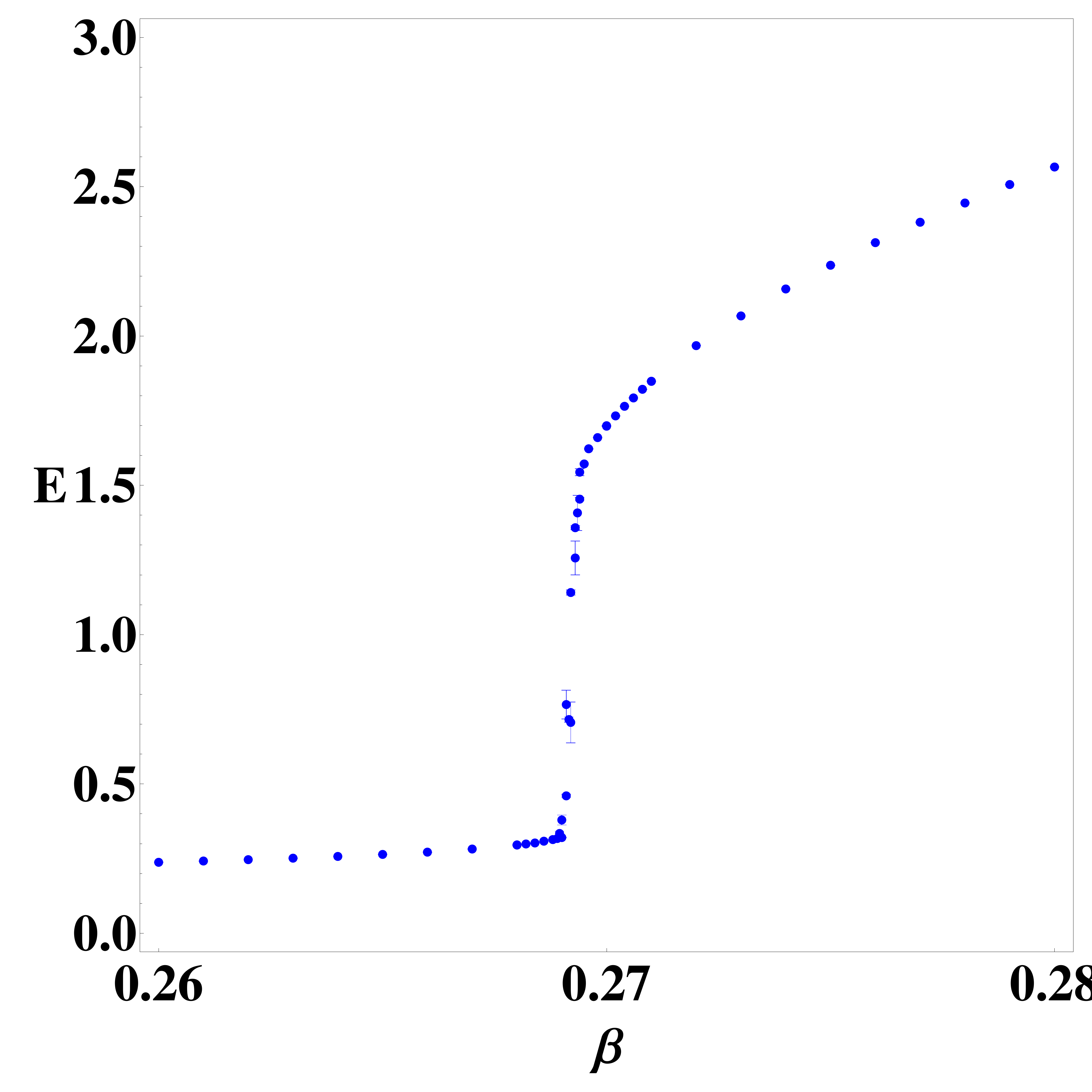}
\includegraphics[scale=0.08]{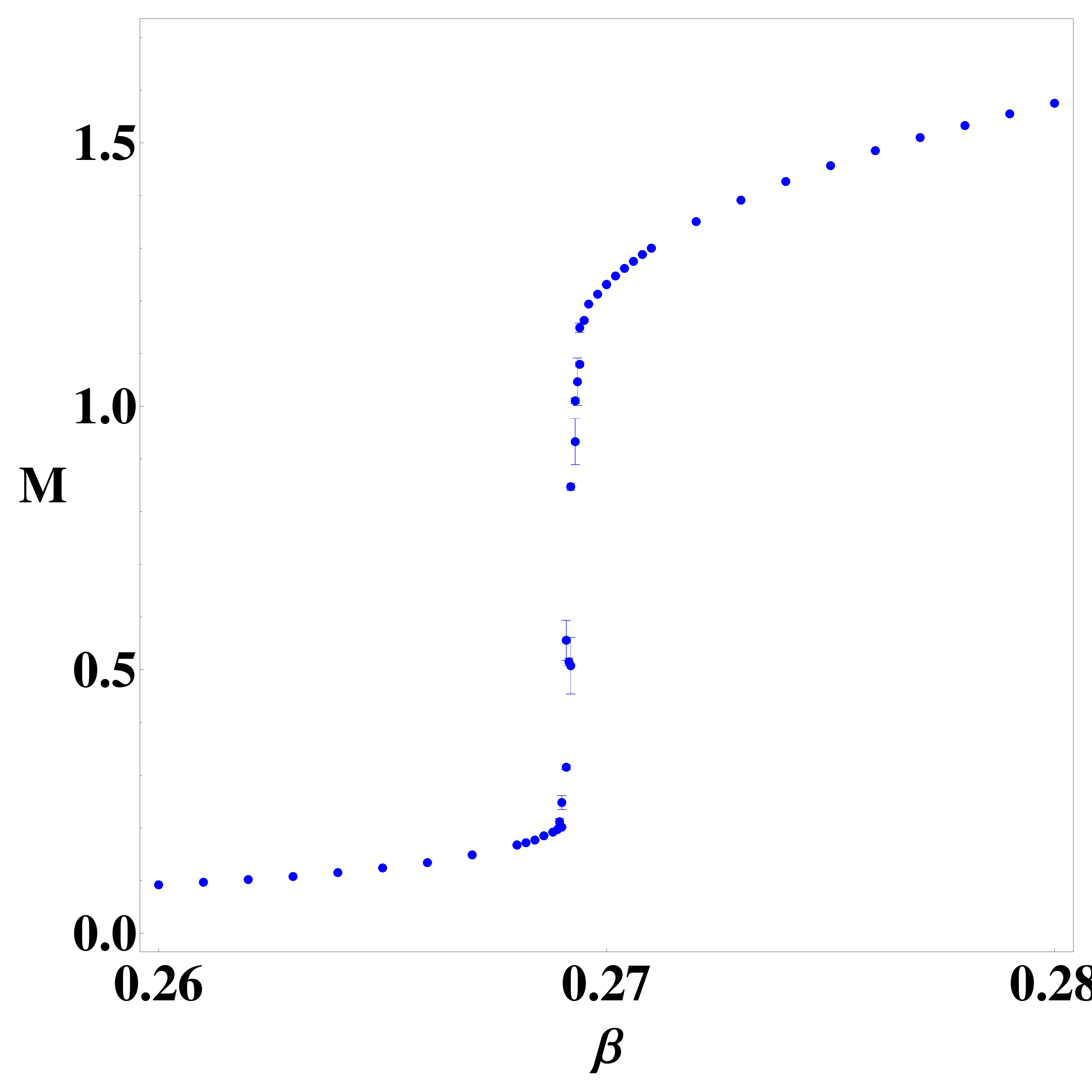} 
\includegraphics[scale=0.08]{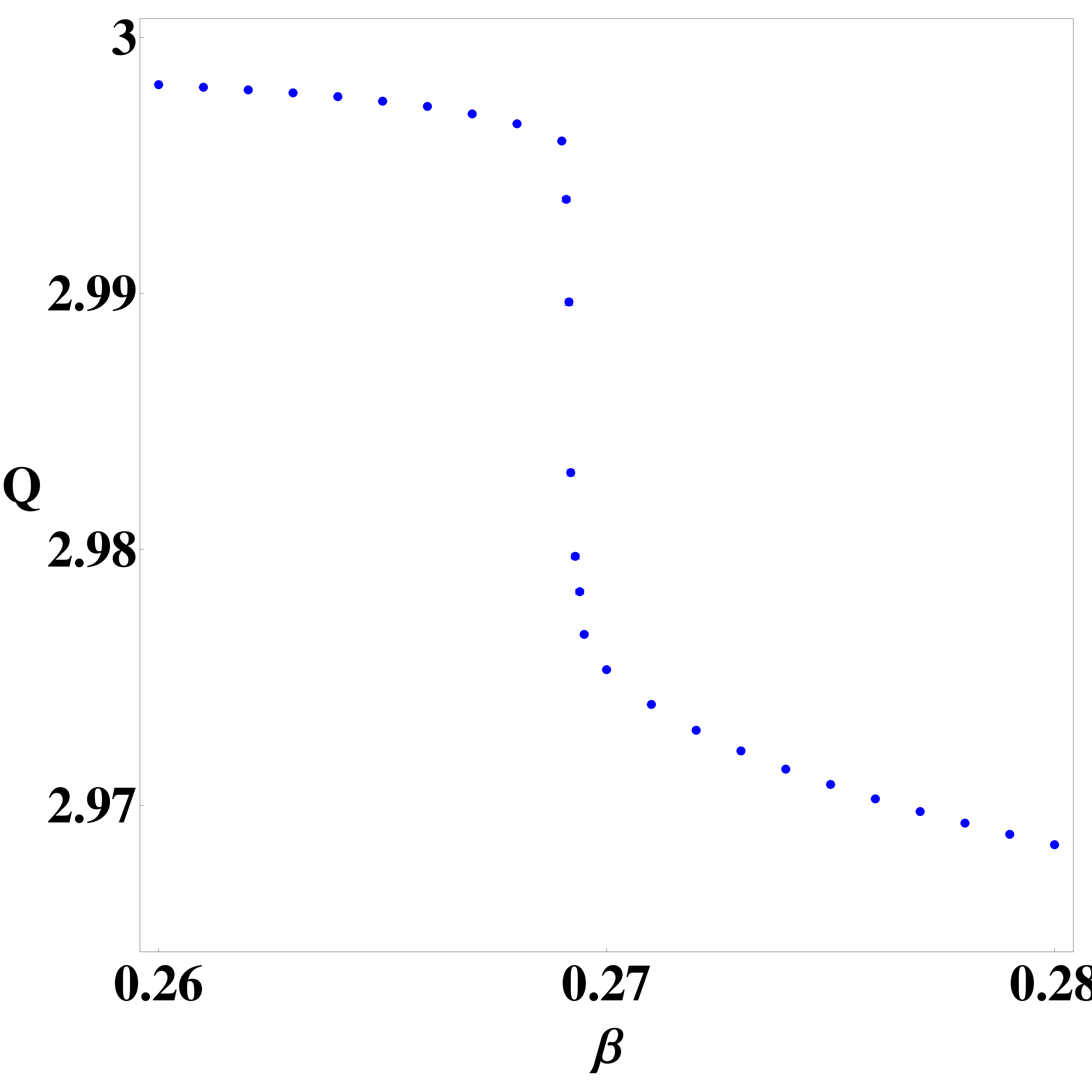}
\caption{Observables for $h=0.01$,  $\mu = 0$ around the first order phase transition: (left to right) energy density, magnetization, quark condensate}
\label{fig:first-order}
\end{figure}

\begin{figure}
\centering
\includegraphics[scale=0.08]{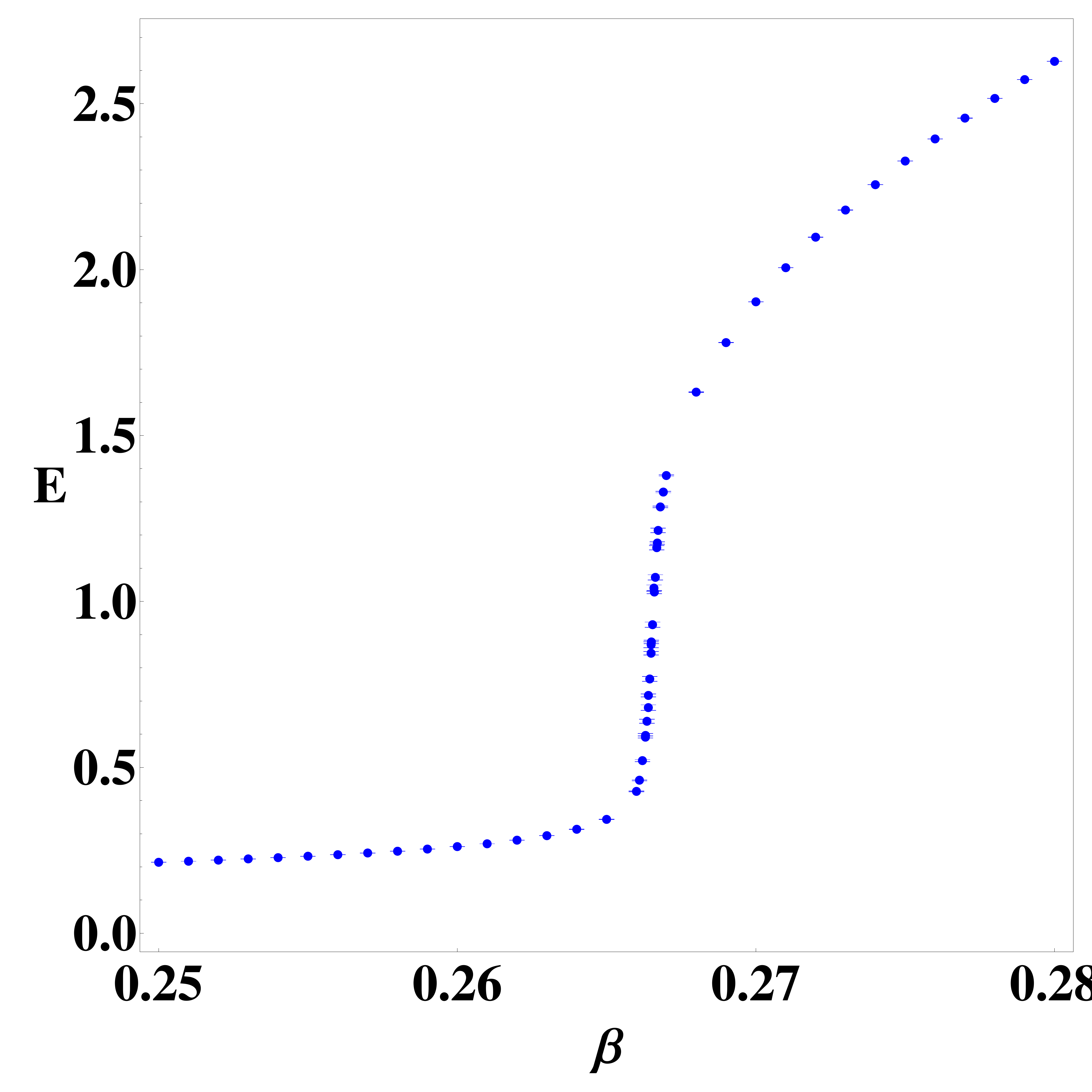}
\includegraphics[scale=0.08]{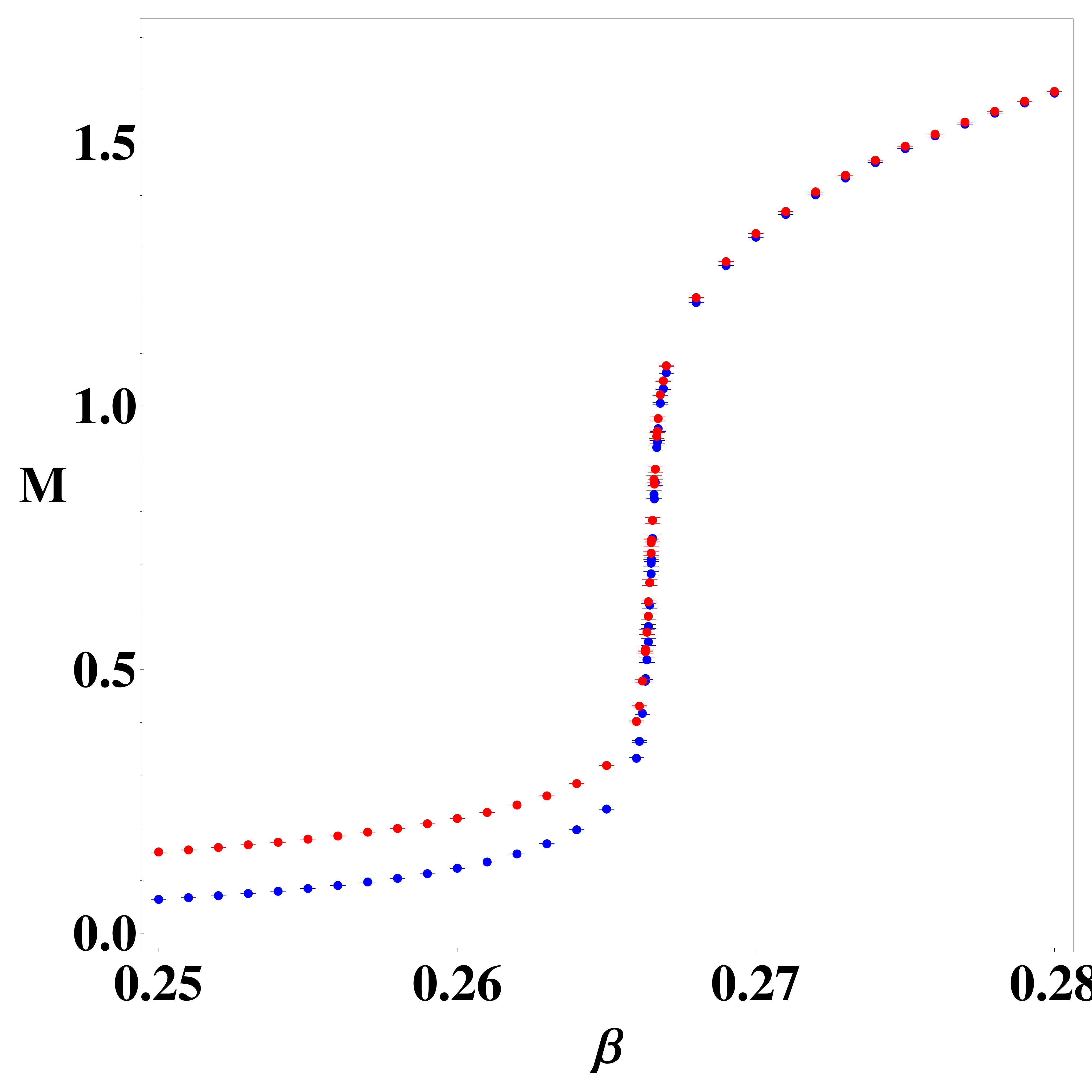} 
\includegraphics[scale=0.08]{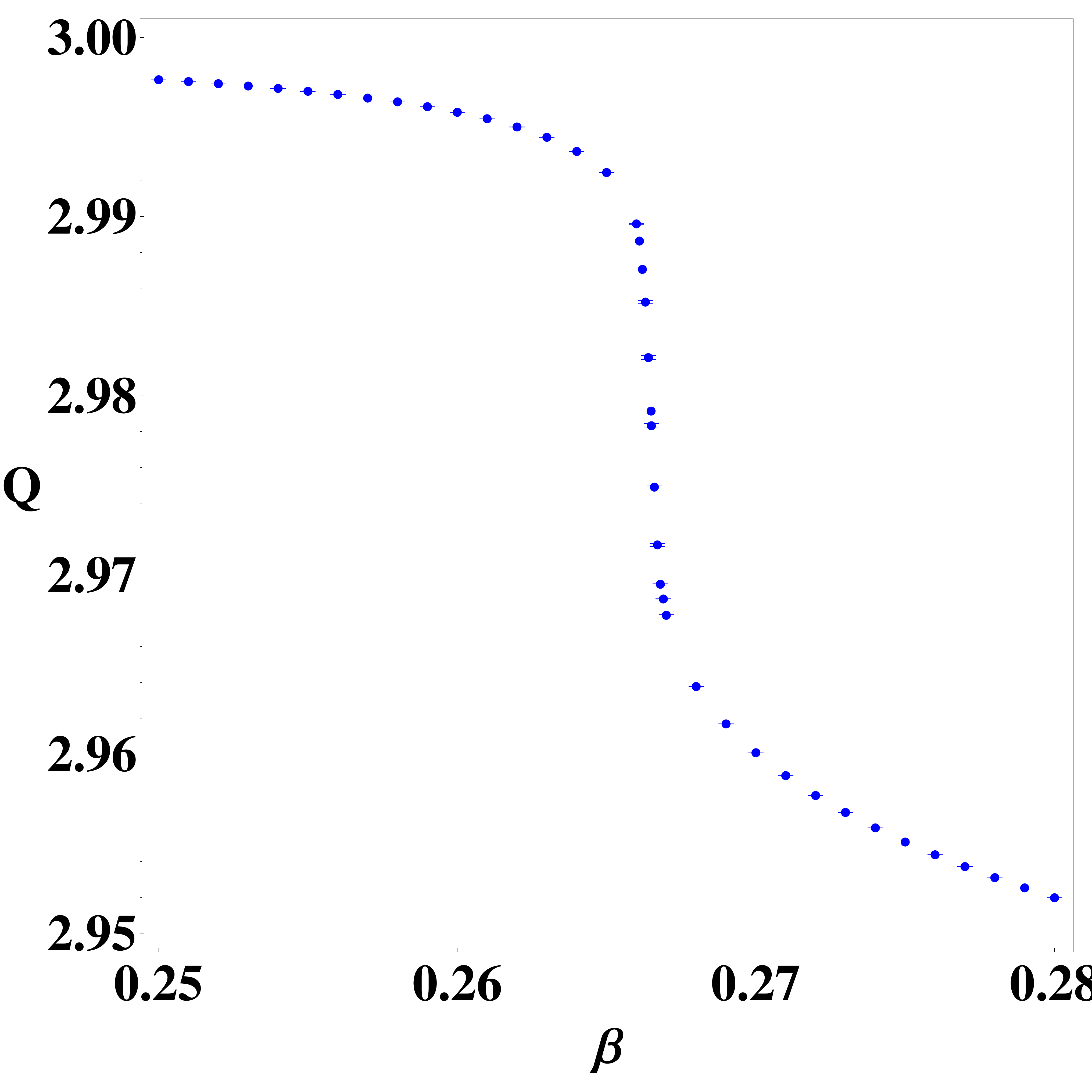}
\includegraphics[scale=0.08]{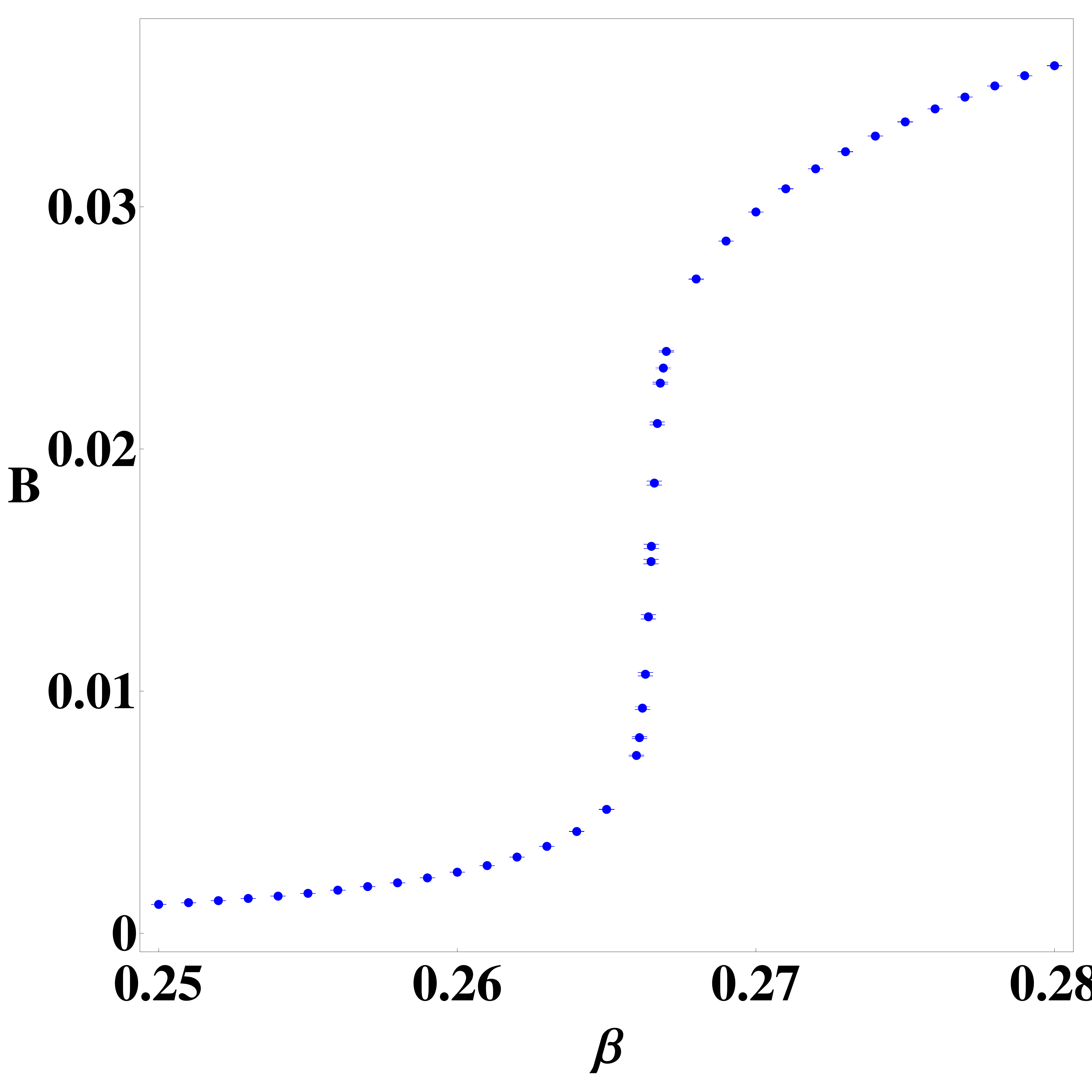}
\caption{Observables for $h=0.01$,  $\mu = 0.9635$ close to the second order phase transition: (left to right) energy density,
 magnetization (blue) and conjugate magnetization (red), quark condensate, baryon number density}
\label{fig:second-order}
\end{figure}

\begin{figure}
\centering
\includegraphics[scale=0.08]{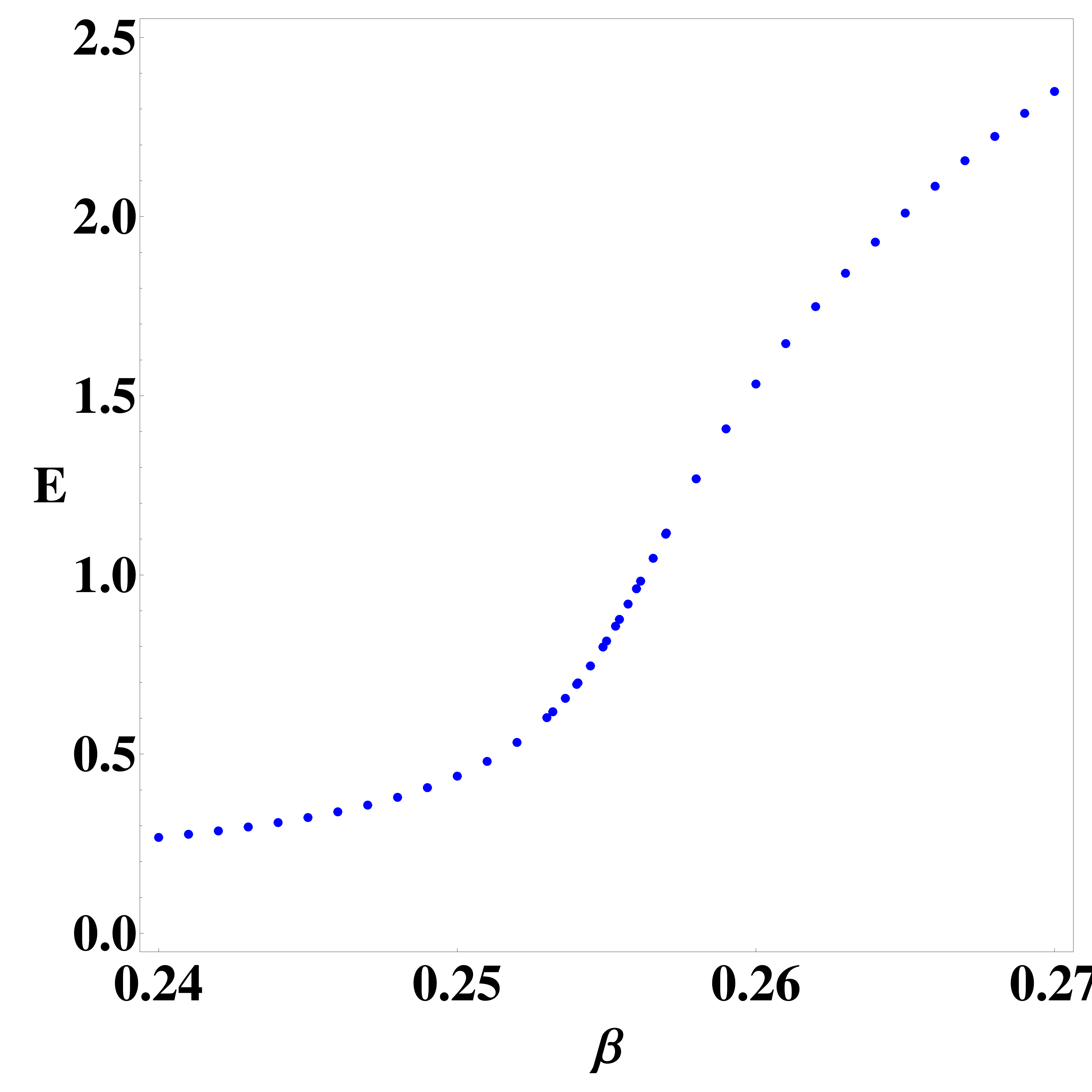}
\includegraphics[scale=0.08]{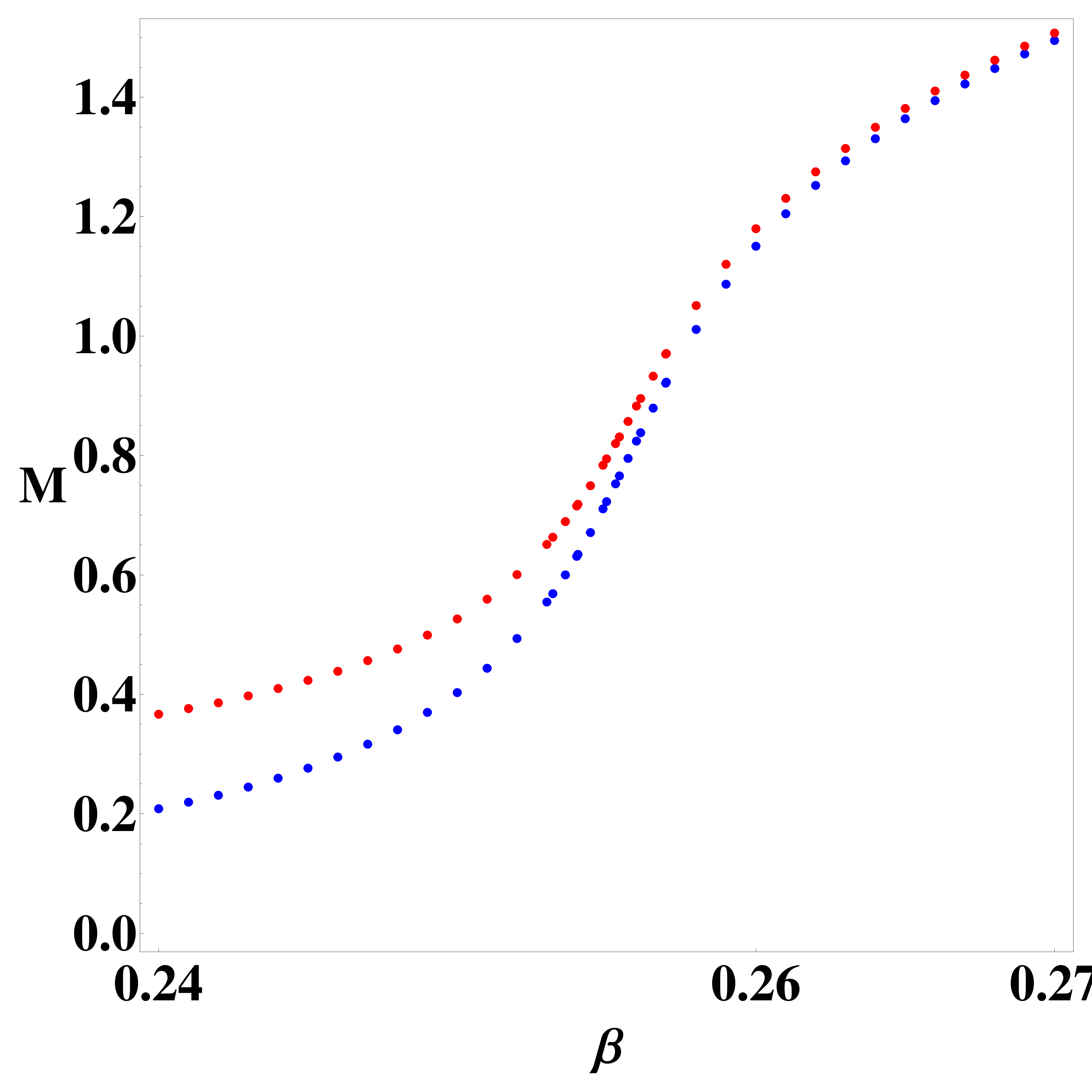} 
\includegraphics[scale=0.08]{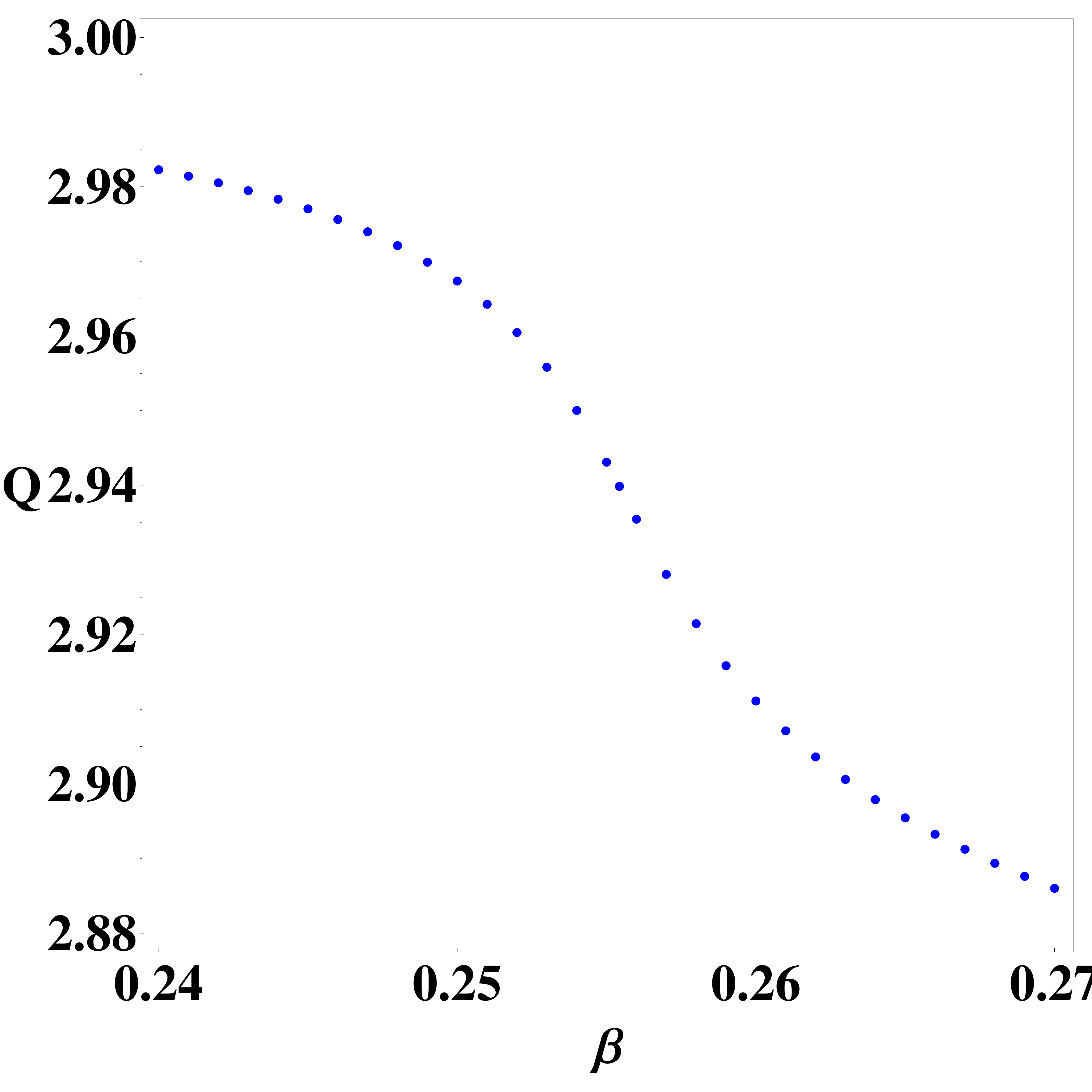}
\includegraphics[scale=0.08]{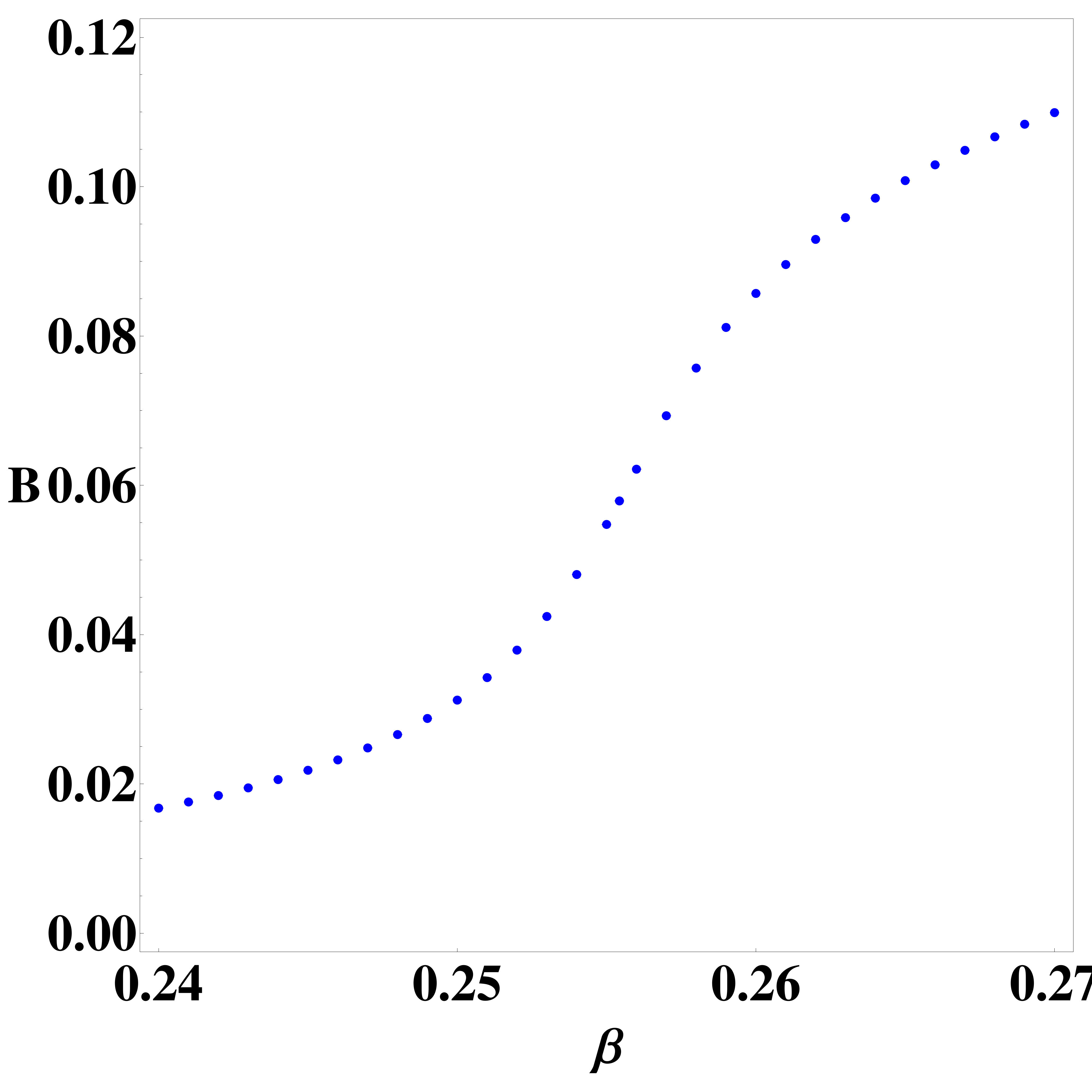}
\caption{Observables for $h=0.01$,  $\mu = 2.0$ in the crossover region: (left to right) energy density,
 magnetization (blue) and conjugate magnetization (red), quark condensate, baryon number density}
\label{fig:crossover}
\end{figure}

The behavior of the observables around the deconfinement phase transition is shown in Figs.~\ref{fig:first-order}-\ref{fig:crossover}. 
All four observables are sensitive to the transition -- for example, for the first order transition, all the observables exhibit a finite jump. One can also notice that while below $\beta_c$ mean magnetization and mean conjugate magnetization are visibly different (while both remaining real), above $\beta_c$ they become much closer, that is, their difference (mean imaginary part of the magnetization) 
goes to zero at large $\beta$. 

\section{Correlation functions and screening masses}

Since we have two components of mean Polyakov loop, 
($\left\langle \TR U \right\rangle$ and 
$\left\langle \TR U^\dagger \right\rangle$, or equivalently
$\left\langle \RE \TR U \right\rangle$ and 
$\left\langle \IM \TR U^\dagger \right\rangle$), 
we have four components of the Polyakov loop correlation function, which we write in the matrix form following~\cite{correlation-matrix}
\begin{align}
\label{corr-matrix}
\Gamma(x,y) &= \begin{pmatrix}
\left\langle \RE \TR U(x) \RE \TR U(y) \right\rangle & 
\left\langle \RE \TR U(x) \IM \TR U(y) \right\rangle \\
\left\langle \IM \TR U(x) \RE \TR U(y) \right\rangle & 
\left\langle \IM \TR U(x) \IM \TR U(y) \right\rangle 
\end{pmatrix} = \ \\
\nonumber
&= \begin{pmatrix}
\Gamma_{rr}(x,y) & 
\Gamma_{ri}(x,y) \\
\Gamma_{ri}(x,y) & 
\Gamma_{ii}(x,y) 
\end{pmatrix} \ .
\end{align}

For $\mu = 0$ the off-diagonal terms are zero, 
and the coefficients in the exponential decay of the connected parts in diagonal terms define the magnetic and electric screening masses,
\begin{align}
\label{em-screening}
\Gamma_{rr, \rm conn}(0, R) &\sim \frac{e^{-m_M R}}{R^\eta} \ , & \Gamma_{ii, \rm conn}(0, R) &\sim \frac{e^{-m_E R}}{R^\eta} \ .
\end{align}

For $\mu > 0$ the electric and magnetic sector mix, so 
each correlation matrix element should be a sum of two terms
-- one decaying with $M_m$, and the other with $M_e$. 

The mean-field analysis following the large-$N$ limit approach of~\cite{thooft-veneziano-limit} gives the following expressions for the correlation functions decay:
\begin{align}
\label{real_real_corr}
\Gamma_{rr, \rm conn}(0, R) &\sim  
 ( \sqrt{C_1} M + \sqrt{C_2} M^*)^2 G_R(m_M)  - ( \sqrt{C_1} M - \sqrt{C_2} M^*)^2 G_R(m_E) \ , \\
\label{im_im_corr}
\Gamma_{ii, \rm conn}(0, R) &\sim
 ( \sqrt{C_1} M + \sqrt{C_2} M^*)^2 G_R(m_E)  - ( \sqrt{C_1} M - \sqrt{C_2} M^*)^2 G_R(m_M) \ , \\
\label{real_im_corr}
\Gamma_{ri, \rm conn}(0, R) &\sim
 ( C_1 M^2 - C_2 M^{*,2}) (G_R(m_M) - G_R(m_E)) . 
\end{align}
Here $G_R(m)$ is the massive Green function (having asymptotic behavior like in (\ref{em-screening})), 
$M$, $M^*$ are mean magnetization and mean conjugate magnetization, $C_1$, $C_2$ are coefficients defined from the Taylor expansion of the action around the saddle point. 

\begin{figure}
\centering
\includegraphics[scale=0.5]{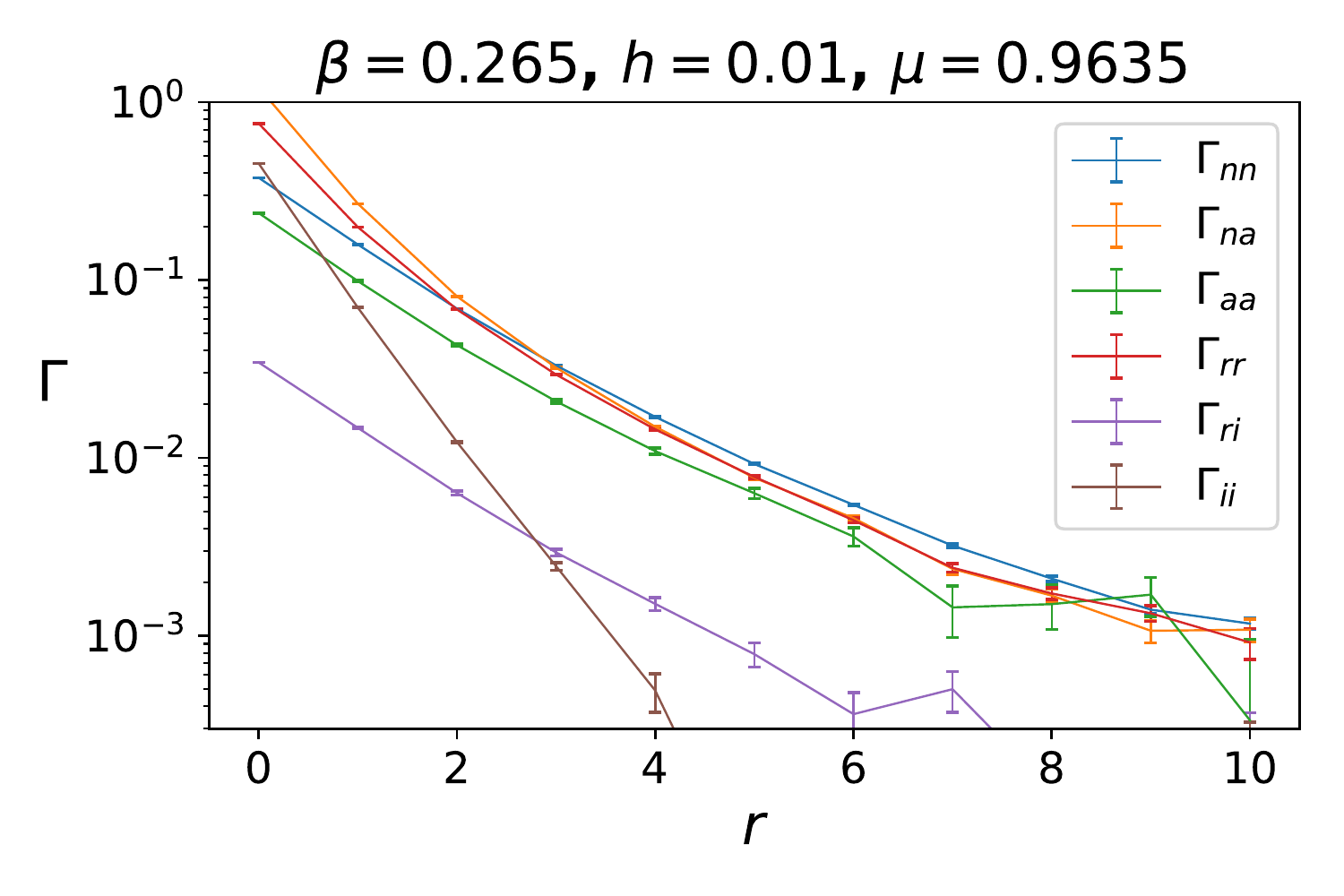}
\caption{Decay of different connected Polyakov loop correlation functions at $\beta=0.265$, $h=0.01$,  $\mu = 0.9635$}
\label{fig:corr-function}
\end{figure}

An example of the Polyakov loop correlation function behavior
obtained from the numeric simulations is given in Fig.~\ref{fig:corr-function}. As we see,
the correlation of real and imaginary parts of the Polyakov loop has the same slope as the correlation of real parts of the Polyakov loop, but is much smaller (the factor before $G_R(m_M)$ in~(\ref{real_im_corr}) is much smaller than the one in~(\ref{real_real_corr})). The slope for the imaginary-
imaginary correlation is larger, which seems to suggest 
that the imaginary-imaginary correlation does not have a term 
with $G_R(m_M)$. Most probably this is a result of very small mixing -- the coefficient before $G_R(m_M)$ in~(\ref{im_im_corr}) is so small that, on the small distances like the one studied here, the slope is defined by $m_E$. 
This would imply that continuing the plot to much larger 
$R$ we would see the imaginary-imaginary correlation finally going parallel to the other ones. Unfortunately, the precision of determination of the correlation functions at large $R$ is not enough to check this prediction.

\begin{figure}
\centering
\includegraphics[scale=0.08]{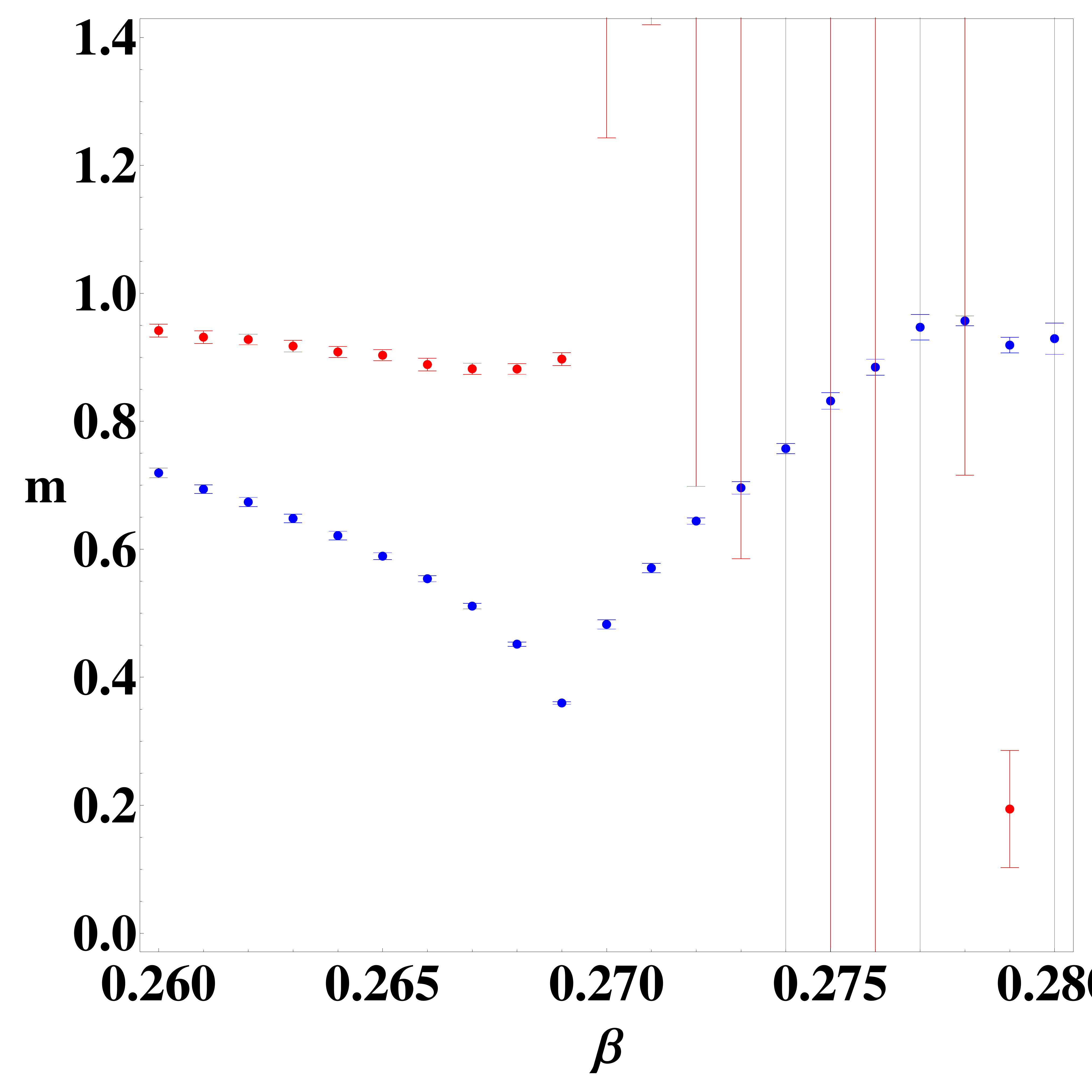}
\includegraphics[scale=0.08]{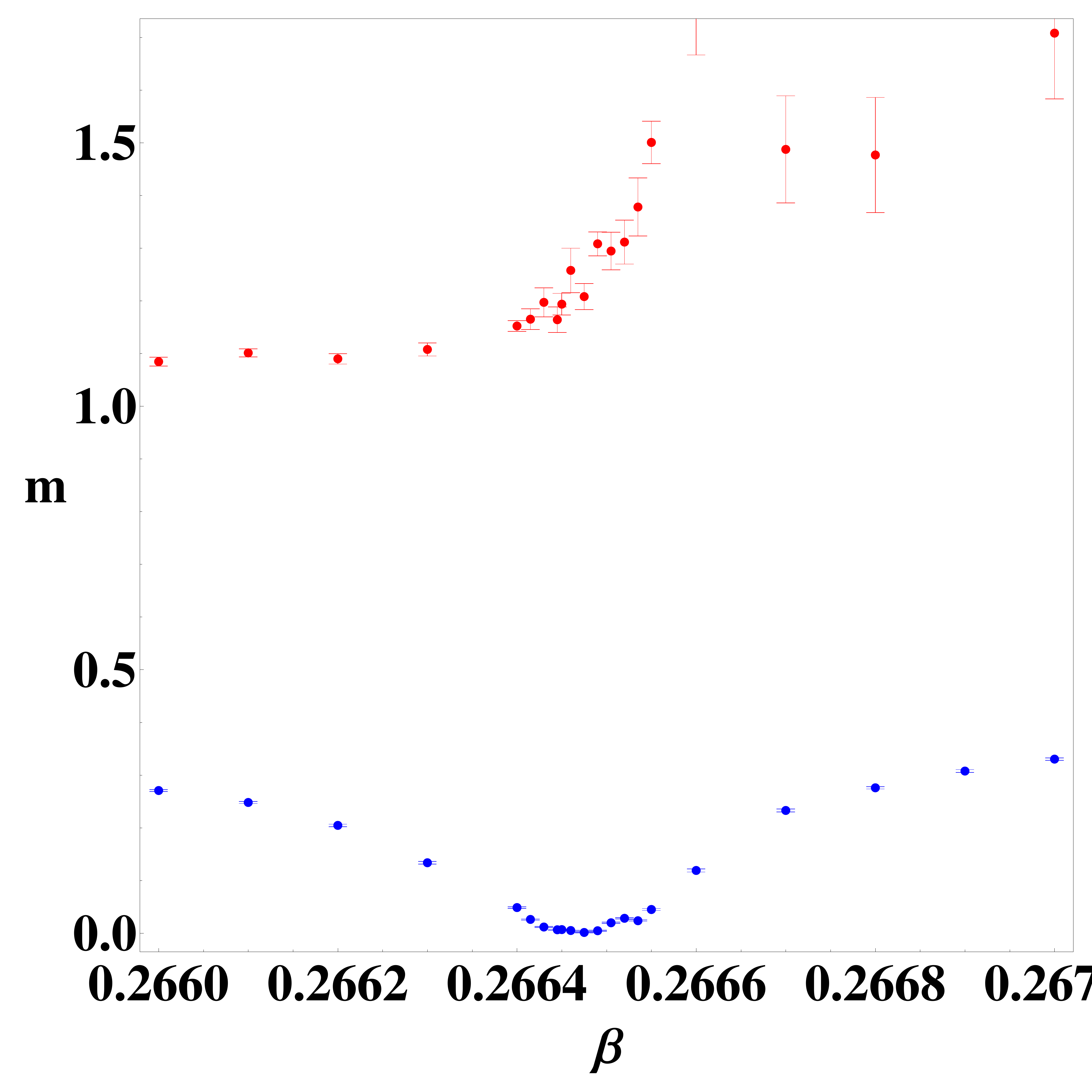}
\includegraphics[scale=0.08]{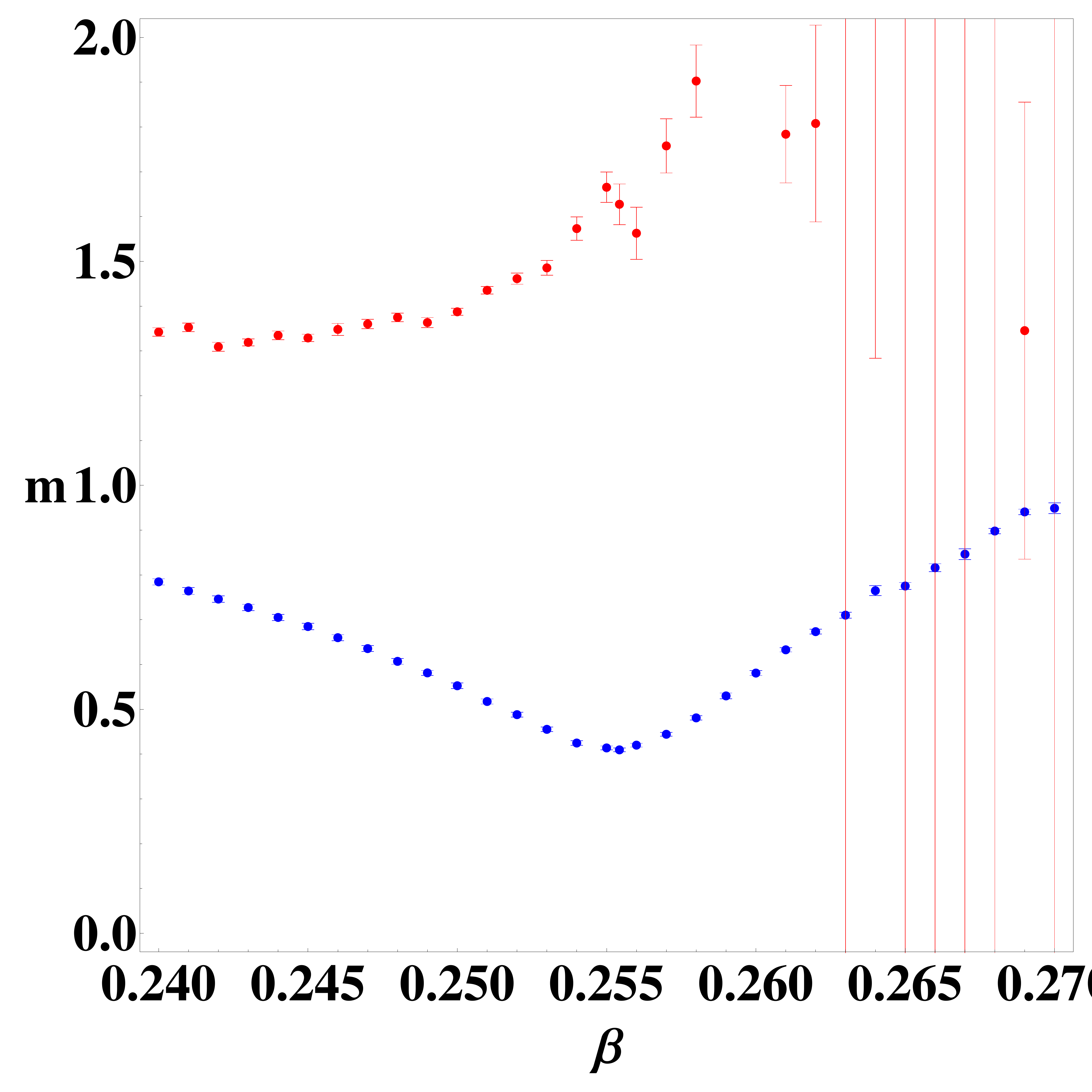}
\caption{Electric and magnetic screening masses at $h=0.01$, $\mu$ = 0 (left), 0.9635 (middle), 2 (right)}
\label{fig:screening-masses}
\end{figure}

Diagonalizing the correlation matrix and extracting the 
masses from a fit of the diagonal components to~(\ref{em-screening}), we obtained the screening masses behavior in different phases (Fig.~\ref{fig:screening-masses}). As can be seen around the transition point the magnetic screening mass exhibits a drop 
(to zero in the case of a second order transition). The electric mass is larger than the magnetic one, and remains almost constant for $\beta < \beta_c$, rising quickly for $\beta > \beta_c$ (the errors grow in this region due to a need to extract the exponential decay with large mass from a set of data with finite absolute precision).

\section {Summary}

 A dual formulation free from sign problem is obtained for an effective Polyakov loop model with an exact static quark determinant. The model can be simply generalized to larger $N$, and $N_f$.
    
 Numeric simulations in a wide parameter region show the location of the second order critical line that separates the region of first order phase transition from the crossover region.
    
 The Polyakov loop correlation function behavior is governed by two masses. Around the phase transition the magnetic (smaller) mass exhibits a rapid drop (to zero for the second order transition), while the electric (larger) mass remains almost constant at $\beta < \beta_c$, and grows rapidly for $\beta > \beta_c$.
    These results are in qualitative agreement with the prediction of mean field from the large-$N$ limit approach.
    
A more detailed description of our results for the phase diagram and local observables can be found in~\cite{local-observables}. A paper with a detailed description of the correlation function and screening mass results is currently in preparation.

\section*{Acknowledgements}

Numerical simulations have been performed on the ReCaS Data Center of INFN-Cosenza.
O. Borisenko acknowledges support from the National
Academy of Sciences of Ukraine in frames of the project
"Meeting new experimental data on proton-proton, proton-nuclei,
nuclei-nuclei interactions at high energies at CERN, BNL,
FERMILAB, GSI for theoretical analysis" (No. 0121U112254).
V.~Chelnokov acknowledges support by the Deutsche Forschungsgemeinschaft 
(DFG, German Research Foundation) through the CRC-TR 211 
‘Strong-interaction matter under extreme conditions’ – project number 315477589 – TRR 211. 
E. Mendicelli was supported by the York University Graduate Fellowship Doctoral - International and by the Carswell Graduate Scholarship.
A.~Papa acknowledges support from Istituto Nazionale di Fisica Nucleare (INFN), through the NPQCD project.

\end{document}